\documentclass[article]{aa}
\usepackage{graphicx}
\usepackage{txfonts}

\begin{document}

\title{IGR J17480-2446: a new class of accreting binaries?} 
\author{
A.~Bonanno\inst{1,2}, V.~Urpin\inst{1,3}
}
\offprints{}
\institute{
           $^{1)}$ INAF, Osservatorio Astrofisico di Catania,
           Via S.Sofia 78, 95123 Catania, Italy \\
           $^{2)}$ INFN, Sezione di Catania, Via S.Sofia 72,
           95123 Catania, Italy \\
           $^{3)}$ A.F.Ioffe Institute of Physics and Technology and
           Isaac Newton Institute of Chile, Branch in St. Petersburg,
           194021 St. Petersburg, Russia
}
\date{\today}

\abstract
{The recent discovery of long-period, 
low magnetic field pulsars in low-mass X-ray binaries (LMXBs) represents
a challenge for the standard evolutionary scenario.   
These pulsars have a  magnetic field strength comparable to that of
millisecond pulsars ($\sim 10^8 - 10^9$ G), but their period
is at least an order of magnitude longer.}
{We discuss the origin of this new class of  pulsars within 
the standard picture of LMXBs formation and apply our results 
to the case of IGR J17480-2446.}    
{The magnetothermal 
evolution of the binary system is studied numerically by taking into account  the effect of different accretion rates
during the Roche-lobe overflow in the framework of the minimal cooling scenario.}
{We show that, in addition to standard millisecond pulsars, 
long-period low magnetic 
field pulsars should also  be expected as a possible
outcome of the binary evolution, depending on the 
strength of the accretion rate during the Roche-lobe overflow.
In particular, we argue  that IGR J17480-2446 belongs to this new  class of objects. }  
{}

          
\maketitle         
          
\section{Introduction}
According to the recycling scenario \citep{bhatta91} the population of 
radio millisecond pulsars and low mass X-ray binaries (LMXBs) share a 
common evolutionary link. A young, magnetized, slowly rotating 
neutron star in a binary system can accrete the gas stripped by the donor 
star spinning up until an old radio millisecond pulsar is formed.
Accreting pulsars in LMXBs generally have spin periods lower than 10 ms,  
but a small number of sources with longer periods has also  been found.  
The recently discovered pulsar IGR J17480-2446 (J17480 in the following)
with the spin period 90.6 ms  belongs to this small group \citep{papitto11}.  
During the time spanned by the observations, its  bolometric  luminosity 
varied within the range $(1.7-6.8) \times  10^{37}$ erg s$^{-1}$,  which 
implies an accretion rate on the order of $(1-4) \times 10^{-9}$ 
$M_{\odot}$ yr$^{-1}$. The presence of pulsations throughout the observations 
allowed the authors to estimate the magnetic field at the neutron star 
surface as $\sim 7 \times 10^8$ G,  a very low value for an accreting 
pulsar with such a long period and high luminosity. 

In the standard model of evolution in a binary system 
\citep{pringle72,Illarionov75}, the neutron star during the accretion 
phase initially spins up until it approaches the so-called spin-up line 
corresponding to the accretion rate. The spin-up line in the magnetic field - spin period (B-P) diagram is determined by corotation at 
the Alfven radius. During the further evolution, a balance between 
spin-up and the rate of field decay is reached, so that the 
neutron star slides down the corresponding spin-up line with a rate that 
is determined by field decay \citep{ba91}. In accordance with this model, 
measurements of the spin period and accretion rate determine the position 
of an accreting pulsar in the B-P diagram and can provide an 
estimate of the magnetic field. In the case of J17480, the spin period 
90.6 ms and the accretion rate $\sim 10^{-9}$  $M_{\odot}$ yr$^{-1}$ imply 
that the magnetic field should be $\sim 10^{10}$ G. The field estimated 
by \cite{papitto11} is much lower and most likely the accreting pulsar 
in this system does not slide down on the corresponding spin-up line. 
To the best of our knowledge, J17480 is the first clear example of 
accreting pulsars with a significant departure from the spin-up line. 

In this paper, we demonstrate that pulsars with long spin periods and low
magnetic fields (like J17480) can be the natural products of evolution of 
the neutron star in LMXBs and millisecond pulsars. We show that 
field decay in a neutron star during the accretion phase can be faster
than spin-up, and a significant departure from the spin-up line can be 
expected during the accretion phase. This mechanism can be the central issue 
in understanding the origin of a particular class of slowly rotating 
accreting pulsars in LMXB.

\section{Evolutionary model}

 We consider the magnetic and spin evolution of a neutron star in a binary 
system with a low-mass companion. We do not specify the mechanism  
responsible for the origin of the magnetic field in a neutron star. We 
only assume that the field is maintained by electric currents in the 
crust. This type of magnetic configuration can be generated, for example, 
by turbulent dynamo at the beginning of a neutron star life 
\citep{bonanno05,bonanno06}.

The binary is assumed to be relatively close, thus the companion can fill 
the Roche lobe at the end of its main-sequence life. During its main-sequence 
evolution (which lasts  $\sim 10^9 - 10^{10}$ yrs), the secondary loses 
mass because of the stellar wind and some fraction of the wind plasma can 
generally be captured and accreted by the neutron star at a certain stage. 
Both the magnetic and spin evolution may be affected by this accretion and, 
as a result, the neutron star passes through different evolutionary 
stages. After the secondary leaves the main sequence and fills its Roche 
lobe, the rate of mass loss and, hence, the accretion rate can be drastically 
enhanced. Detailed discussion of evolutionary transformations of the neutron 
star in a binary system is given by \cite{urpin98mn} and we follow this 
scenario. We note that the developed evolutionary model  successfully explains 
   the origin of millisecond pulsars \citep{urpin98aa}, the 
magneto-rotational evolution of the pulsars in high mass binary systems 
\citep{urpin98hm}, and the Her X-1 like pulsars \citep{konenkov98}. 

The most important point in understanding the neutron star evolution is the behavior of 
the crustal magnetic field.
This  is determined mainly by the conductive 
properties of the crust and the material motion throughout it. In a very 
strong magnetic field, the ohmic  dissipation can be accompanied by 
non-dissipative Hall currents which affect the field decay indirectly, 
coupling different modes and redistributing the energy among modes. We note 
that because of the nonlinear nature of the Hall currents, they can 
generate magnetic features with a smaller length-scale than the background 
magnetic field \citep{naito94}. However, these currents are important only 
at the very early stage of evolution and only if the initial field is very 
strong. We are interested in the long term evolution and we will neglect 
the Hall currents. Then, the induction equation in the crust reads 
\begin{equation}
\frac{\partial \vec{B}}{\partial t} =  - \frac{c^2}{4 \pi} \nabla
\times \left( \frac{1}{\sigma} \nabla \times \vec{B} \right) + 
\nabla \times ( \vec{v} \times \vec{B}),
\end{equation}
where $\sigma$ is the conductivity and $\vec{v}$ is the velocity of
crustal matter. The velocity $\vec{v}$ is caused by the flux of the 
accreted matter and it is nonzero only during the accretion phases.
Assuming the spherical symmetry of a flow, the velocity in the negative 
radial direction can be estimated as $v = \dot{M}/4 \pi r^2 \rho$, 
where $\dot{M}$ is the accretion rate and $\rho$ is the density. 

We consider the evolution of an axisymmetric field. Following 
\cite{wendell87} we introduce the vector potential $\vec{A} = (0, 
0, A_{\varphi})$, where $A_{\varphi} =  S(r, \theta, t)/r$ and $(r, 
\theta, \varphi)$ are the spherical coordinates. In the case of a 
dipole field, $S$ can be represented as $S = s(r, t) \sin \theta$. 
Then, we have 
\begin{equation}
\frac{\partial s}{\partial t} = v \frac{\partial s}{\partial r} +
\frac{c^2}{4 \pi \sigma} \left( \frac{\partial^2 s}{\partial r^2} -
\frac{2}{r^2} s \right).
\label{induction}
\end{equation}
The function $s(r, t)$ is related to the surface magnetic field at 
the pole, $B_p(t)$, by $B_p(t) = 2 s(R, t)/ R^2$ \citep{urpin94}. 
We normalize $s$ in such a way that $s(R, 0)=1$. Then, the ratio 
$B_p(t)/ B_p(0)$ is given by $s(R, t)$. Continuity of $\vec{B}$ at 
the surface $r=R$ yields the boundary condition $\partial s / 
\partial r + s / R= 0.$ Since $\sigma$ depends on the temperature 
$T$, accretion can influence the field decay by changing the 
temperature $T$. Therefore, the magnetic evolution of a neutron star 
is coupled essentially to its thermal history. 
This is in contrast to the widely accepted prescription of the
magnetic field decay stating that the decay is proportional to the
amount of accreted material \citep{shiba}.
However, such a simple model is obviously insufficient from a theoretical
point of view because the decay is determined not only by the duration
of accretion (or, equivalently, by the total amount of accreted mass), 
but also by the conductivity of the crust. The latter
depends on the temperature, which is also dependent on the accretion rate.  

Therefore, a consistent description of the field decay 
must take into account  both the duration of accretion (or
total amount of accreted mass) and accretion rate. In particular, a simple model 
with a field decay proportional to the amount of accreted material
is not in agreement with some observational data. This point is discussed
in detail by \cite{wije}. In the model developed in this paper,
the magnetic field decay is determined by both the accretion rate and 
duration of accretion.


During its evolution, the neutron star can be processed in four 
main evolutionary phases:

\noindent
{\it Phase (i) }. During the initial phase, the neutron star is not 
affected by its companion because the pressure of magnetodipole waves 
emitted by a rapidly rotating neutron star is very high and prevents 
the wind plasma of the companion from interaction with the magnetosphere. 
Thus, the magnetic, thermal, and spin evolution follow those of an 
isolated star. The magnetic field is governed by Eq. (\ref{induction}) 
with $v=0$. Since the wind plasma does not interact with the neutron 
star magnetosphere, the spin evolution is determined by magnetodipole 
radiation,
\begin{equation}
P \dot{P} = \frac{2 \pi^2 B_p^2 R^6}{3 c^3 I},
\end{equation}
where $P$ is the spin period and $I$ is the moment of inertia.  Phase 
{\it (i)} lasts as long as  the wind plasma is stopped by the pressure 
of magnetodipole radiation behind the radius of gravitational capture. 
The stopping radius, $R_s$, is determined by the balance between the 
dynamical pressure of the wind and the pressure of magnetodipole waves. 
The pressure of the wind is $\sim \rho_w V_w^2/2$, where $\rho_w$ and 
$V_w$ are the density and velocity of the wind plasma, respectively. The 
density of the wind at the distance $r$ from the secondary may be 
estimated as $\rho_w \sim \dot{M}_0 / 4 \pi r^2 V_w$ with $\dot{M}_0$ 
being the rate of mass loss of the secondary. The pressure of the 
magnetodipole radiation at the distance $R_s$ from the neutron star is 
$\sim (1/4 \pi R_s^2) (B_p^2 R^6 \Omega^4/6 c^4)$, where $\Omega$ is the 
angular velocity of a neutron star. Equating these pressures, we obtain 
\begin{equation}
R_s = a \left[ 1 + \left(\frac{3 c^4 \dot{M}_0 V_w}{B_p^2 R^6 \Omega^4}
\right)^{1/2} \right]^{-1},
\end{equation}
where $a$ is the separation between stars. If $R_s$ 
is greater than the radius of gravitational capture $R_G =2 G M/ V_w^2$  
\citep{bondi52}, being $M$  the neutron star mass, the wind matter is not 
captured by the neutron star and it cannot interact with the magnetosphere.  
The condition $R_s = R_G$ thus determines the end of  phase {\it (i)}. 

\noindent
{\it Phase (ii)}. If $R_G > R_s$, a fraction of the wind plasma starts 
to be captured by the neutron star. The rate of gravitational capture of 
the wind plasma, $\dot{M}$, is approximately equal to $\dot{M} = \dot{M}_0 
(R_G/ 2 a)^2$ if $a > R_G$. Only this fraction of the wind can directly 
interact with the magnetosphere. Rotation of the neutron star and its 
magnetosphere is rather fast at this phase, and the magnetosphere acts as 
a propeller, ejecting the wind plasma \citep{Illarionov75}. Since accretion 
is forbidden during this phase, the thermal and magnetic evolution does 
not differ from that of an isolated neutron star. The spin evolution can 
be different, however, because a fraction of the stellar angular 
momentum is transferred to the wind plasma which penetrates into the 
magnetosphere. It is usually assumed  that the wind plasma interacts with 
the magnetosphere at the Alfven radius, $R_A$, which is determined by a 
balance of the Lorentz force and the dynamical pressure of plasma,
\begin{equation}
R_A = \left( \frac{2 R^6 B_p^2}{\dot{M} \sqrt{GM}} \right)^{2/7}.
\end{equation}  
If the angular velocity of the neutron star is larger than the Keplerian 
angular velocity at the Alfven radius, $\Omega_K(R_A) = (G M / R_A^3)^{1/2}$, 
then the wind plasma penetrating to the Alfven radius can extract
some portion of the angular momentum from the rapidly rotating magnetosphere. 
The rate of the angular momentum loss by the neutron star, $\dot{J}_p$, can 
be estimated as $\sim (\Omega_K(R_A) R_A^2)\dot{M}$. Then, the corresponding 
spin-down rate of the neutron star during the propeller phase is
\begin{equation}
\dot{P} \sim P^2 \dot{J}_p / 2 \pi I \approx \beta P^2 B_p^{2/7}
\dot{M}  ^{6/7},
\end{equation}  
where $\beta = (G M R^2/ 4)^{3/7}/ \pi I$. Rotation slows down until 
the angular velocity of the neutron star becomes comparable to 
$\Omega_K(R_A)$. This condition determines the critical period, $P_{eq}$, 
at which the propeller phase ends,
\begin{equation}
P_{eq} = \frac{2 \pi}{(GM)^{5/7}} \left( \frac{R^6 B_p^2}{4 \dot{M}}
\right)^{3/7}  \approx \frac{12.4 B_{12}^{6/7} R_{6}^{18/7}}{\dot{M}_{-10}^{3/7}} 
\; {\rm s},
\end{equation}
where $B_{12}= B_p/10^{12}$ G, $\dot{M}_{-10} = \dot{M} /10^{-10}
M_{\odot}$yr$^{-1}$, and $R_{6}= R/10$ km. This equation determines the
 spin-up line in the $B-P$ plane.

\noindent
{\it Phase (iii)}. This phase begins when $\Omega$ becomes smaller than 
the Keplerian angular velocity at the Alfven radius. The neutron star can no longer 
work as a propeller,  and the accretion from the wind is allowed.
Nuclear burning of the accreted material heats the crust, decreases the crustal
conductivity, and can induce a significant accretion-driven field decay.
Therefore, the magnetic evolution during  phase {\it (iii)} is driven by 
Eq. (\ref{induction}) with $v \neq 0$. The thermal evolution is determined 
mainly by pycnonuclear fusion in the crust. The thermal structure of the star 
with pycnonuclear reactions has  been studied by a number of  authors 
\citep{hae90,zdu92}.

The rotational evolution differs from that during  phases {\it (i)} and 
{\it (ii)}. The accreting matter very likely carries some amount of the angular 
momentum and transfers it to the neutron star. The angular momentum 
carried by the wind plasma can be characterized by its Keplerian 
value, $\rho_w \Omega_K(R_A) R_A^2$, multiplied by an efficiency 
factor, $\xi < 1$. This factor depends very strongly on the accretion flow. 
For example, if the accreted matter forms an accretion disk around the 
neutron star then $\xi \sim 1$. If the disk is not formed (which is more 
typical for  wind accretion), the factor $\xi$ is much smaller ($\sim 
0.1-0.01$). The rate of the angular momentum transfer from the wind to 
the neutron star is $\sim \xi 4 \pi R_A^4  \rho_w \Omega_K v_r$. Then, 
the corresponding spin up rate is
\begin{equation}
\dot{P} \sim - \xi \beta P^2 B_p^{2/7} \dot{M}^{6/7}.
\label{otto}
\end{equation}
As a result of this angular momentum transfer, the neutron star can spin up to 
a shorter period, which is  always larger that the Keplerian 
period at the Alfven radius, otherwise the star would work as a propeller.
\cite{ba91} argued that the neutron star slides down the corresponding 
spin-up line with a rate determined by field decay. This evolution can 
last  either until the accretion regime is changed (for instance, because 
the companion fills its Roche lobe and an enhanced accretion phase starts) 
or until the magnetic field becomes too weak to maintain a balance in spin up 
and the rate of the field decay. In the latter case, the neutron star 
leaves the spin up line and evolves in a very particular way. In our 
calculations, we assume that the spin period of a neutron star follows 
Eq.~(\ref{otto}) if $P \geq P_{eq}$. However, according to the model 
above discussed, if Eq.~(\ref{otto}) leads to $P < P_{eq}$, we 
then suppose $P = P_{eq}$. 

\noindent
{\it Phase (iv)}. After the end of the main-sequence evolution, the
secondary star fills the Roche lobe and the accretion is substantially
enhanced. The neutron star is heated to much higher temperatures and 
the field decay is substantially faster than in the previous phase. In 
LMXBs, accretion due to Roche-lobe overflow lasts as long as $10^6 - 10^7$ 
yrs, and the magnetic field can be drastically reduced 
The field 
decay is still governed by Eq.(\ref{induction}) with $v \neq 0$. The 
thermal structure of accreting neutron stars has been studied by many 
authors \citep{fuji84,mira90,brown98}.  The accreted matter 
probably forms the Keplerian disk during  phase {\it (iv)}. Therefore, spin up is 
driven by Eq. (\ref{otto}) with $\xi \approx 1$. The neutron star spins up 
in accordance with Eq. (\ref{otto}) until it reaches the spin-up line 
corresponding to an enhanced accretion rate. During the further  evolution, 
a balance has to be reached between spin up and the rate of the field 
decay as during phase {\it (iii)}, and  the neutron star slides 
down the spin up line with a rate that is determined by the new, much 
faster, field decay.  The star can leave the spin-up line if the 
field becomes too weak to maintain a balance between spin up and field 
decay. 

When accretion is exhausted, we do not follow  the evolution further. 
Usually, the magnetic field of a pulsar turns out to be rather weak 
after all these evolutionary transformations and its final location 
in B-P plane does not change significantly. 

\section{Numerical results}

In our calculations we made use of the public code developed by Dany Page, 
{\em NSCool}\footnote{\url{http://www.astroscu.unam.mx/neutrones/NSCool/}}
in which  the magnetothermal evolution has been taken into account by 
coupling the induction equation with the thermal equation via the Joule 
heating as discussed in \cite{boba}. In particular, the induction 
equation was solved via an implicit scheme and the electron 
conductivity was calculated with the approach described in 
\cite{pote99} \footnote{\url{http://ioffe.ru/astro/conduct}}. 
Actual calculations were performed for a neutron star of $M = 1.4 
M_{\odot}$ based on the APR EOS \citep{apr} within the 
minimal cooling framework \citep{page04}. The radius and the thickness 
of the crust for this star are 11.5 km and 1.0 km, respectively. For the 
crustal composition, we use the so-called accreted matter model by \cite{zdu08}. 
 
The impurity parameter, $Q$, is assumed to be constant throughout the
crust and equal to 0.001. The initial spin period is assumed to be 
$P_0 = 0.01$ s, but the results are not sensitive to 
this value. On the contrary, the evolution is strongly dependent on the 
original magnetic configuration. The initial magnetic 
field is assumed to be confined to the outer layers of the crust with 
densities $\rho \leq \rho_0$. The calculations presented here are 
performed for $\rho_0 = 10^{13}$  g/cm$^3$. The initial field strength 
at the magnetic pole, $B_p(0)$, is $6 \cdot 10^{12}$ and $2 \cdot 10^{13}$ G. 
The main-sequence lifetime of the companion is as long as $2 \cdot 10^9$
or $6 \cdot 10^{9}$ years. In calculations,  we suppose $\dot{M} 
= 10^{-14}-10^{-15} $ $M_{\odot}$ yr$^{-1}$ during both the propeller and wind 
accretion 
phases: $V_w = 500$ km/s; $a = 5 \times 10^{12}$ cm. After the secondary 
fills its Roche lobe, accretion has to be strongly enhanced and $\dot{M}$ 
can reach the value $5 \cdot 10^{-9}$ $M_{\odot}$ yr$^{-1}$. 

\begin{figure}
\includegraphics[width=9.0cm]{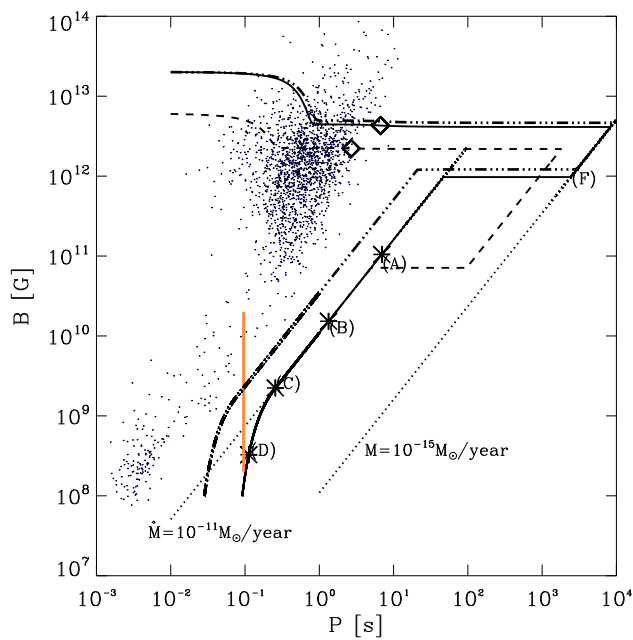}
\caption{Solid line: evolutionary tracks of pulsars in LMXBs with a relatively low
accretion rate during phase IV, $\dot{M} = 10^{-11}$ $M_{\odot}$ yr$^{-1}$,
a wind accretion rate during phase III of $\dot{M} = 10^{-15}$ $M_{\odot}$
yr$^{-1}$,
and an initial magnetic field of $3\cdot10^{13}$ G. 
The end of phase I is marked with a diamond. 
The main-sequence phase ends at point (F)  after $4.5\cdot10^{9}$ years
and marks (A) to (C) represent intervals of 10 Myr after the end of the main sequence.
The total duration of the fast accretion phase is $4.62\cdot10^{7}$ years.
Dashed line: same as dotted line but for an initial magnetic field of $6\cdot10^{12}$ G
and a wind accretion rate of $\dot{M} = 10^{-14}$ $M_{\odot}$ yr$^{-1}$.
Dot-dashed line: same as solid line but 
with a wind accretion rate of $\dot{M} = 10^{-10}$ $M_{\odot}$ yr$^{-1}$. 
The vertical line (in orange) is the observational constraint on the 
magnetic field and period.}
\end{figure}

In Fig.1, we plot the evolutionary tracks for models with a relatively
low accretion rate during  phase (iv), $10^{-10} \geq \dot{M} \geq 10^{-11}$ 
$M_{\odot}$ yr$^{-1}$. We choose such a low accretion rate because 
IGR J17489-2445
is a transient X-ray source. During the several bursts, the accretion rate 
estimated from the luminosity observations is quite high and can reach 
$\sim 10^{-9}$ M$_{\odot}$ yr${-1}$. The accretion rate between 
the bursts
is essentially lower. Since the magnetic evolution of a neutron star
is relatively slow, it is determined  by the average accretion 
rate rather than its variations, which is why we assume in our calculations 
that $\dot{M}$ is lower than the peak value.
In the case of the solid lines,  phase (i)   lasts $5.3 \cdot 10^7$ yrs when the 
neutron star is not influenced by the companion. 
During this phase the field does not decay significantly, thus its surface 
strength is $\approx 10^{12}$ G when the star enters the propeller stage 
(phase (ii)). The duration of this stage is $\sim 7\cdot10^7$ yrs, but the 
field does not decay much either because the star is rather cool and the 
crustal conductivity is high. The wind accretion can last 
sufficiently long, depending on the duration of the main-sequence life. 
Accretion heats the neutron star decreasing the conductivity and slightly 
accelerating field decay. During  phase (iv), accretion is greatly 
enhanced compared to  phase (iii) and the field decay is much 
faster. At some stage, the accreted matter cannot provide a sufficient 
amount of  angular momentum to the neutron star to maintain a balance 
on the spin-up line. Therefore, the neutron star departs from the spin-up line for the considered accretion rates and becomes a long-period 
accreting pulsar. The magnetic field of such a pulsar should be rather 
low because of a fast decay and comparable to the field of millisecond
pulsars. For instance, the magnetic field of the models considered in 
Fig.1 reaches the value $\sim 10^8$ G after $\approx 1.5 \cdot 10^7$ yrs 
of  enhanced accretion. 
However, the periods turn out essentially different for the models 
with the accretion rates $10^{-10}$ and $10^{-11}$ $M_{\odot}$ yr$^{-1}$ with a
longer period corresponding to a low accretion rate. 
Therefore, long-period low magnetized pulsars seem to be a 
natural product of the evolution of neutron stars in LMXBs if the
accretion rate during the Roche-lobe overflow is not very high.

\begin{figure}
\includegraphics[width=9.0cm]{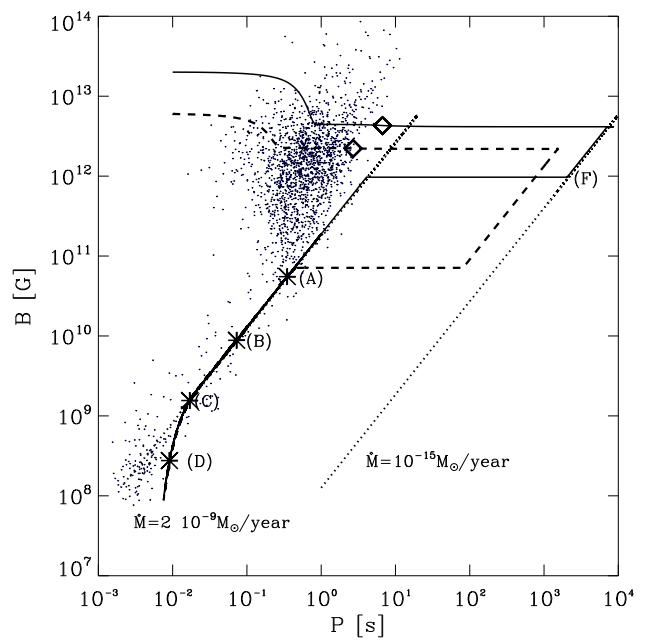}
\caption{Same as Fig. 1 but with an accretion rate during phase IV of 
$\dot{M} = 2 \cdot 10^{-9}$ $M_{\odot}$ yr$^{-1}$. 
The marks (A) to (C) represent intervals of 1 Myr after the end of the 
main sequence. }
\end{figure}

In Fig. 2, we plot tracks of the neutron star processed in all
transformations in LMXBs with a higher accretion rate due to Roche-lobe 
overflow, $\dot{M} = 2 \cdot10^{-9}$ $M_{\odot}$ yr$^{-1}$. The evolution 
is qualitatively the same until the enhanced accretion starts. 
The further evolution differs, however, although the essential changes 
are experienced by the neutron star during  phase (iv), as in  
the previous case. Accretion with the rate $2\cdot10^{-9}$ $M_{\odot}$ yr$^{-1}$ 
heats the neutron star to a temperature $\sim 3\cdot10^{8}$ K and decreases 
the crustal conductivity substantially. Becasue of this, the field is 
reduced by a factor of $\sim 10^3$ after $2.5 \cdot 10^6$ yrs of  
enhanced accretion. The neutron star can maintain a balance on the 
spin-up line much longer than in the case of a weaker accretion. 
However, if accretion lasts sufficiently long, the neutron star 
leaves the spin-up line since the field decay is fast at the end of 
 phase (iv) in comparison with spin-up. Transfer of the 
angular momentum becomes too slow in a weak magnetic field for the following reason. If the accretion flow forms the Keplerian 
disk around a neutron star,
the rate of angular momentum transfer can be estimated as $\dot{M}
\Omega_K(R_A) R_A^2$, where $R_A$ is the Alfven radius that represents
the inner boundary of the disk. We note that this estimate is valid only
if $R_A \gg R$. For the considered accretion rate, this condition is
satisfied if the magnetic field of a neutron star is greater than 
$\sim 10^7$ G. If the magnetic field is very weak ($B_p < 10^7$ G) then
$R_A \leq R$ and the rate of angular momentum transfer is on the order 
of $\dot{M} \Omega_K(R) R^2$. Since $\Omega_K \propto r^{-3/2}$, the
angular momentum transport slows down with a decrease of the magnetic
field and becomes very slow in the case of a weak magnetic field.
In this paper, we do not consider the evolution of neutron stars with 
such a weak field.
We note that the late evolution in all the other cases in Fig. 2 is 
basically the same. As a result, the neutron star after the enhanced 
accretion can have parameters close to those of millisecond pulsars. 
It turns out that the only difference between the neutron stars shown in 
Figs. 1 and 2 is the accretion rate during  phase (iv).

\section{Conclusion}

We have considered the evolution of neutron stars in binary systems with
a low-mass companion. During the course of this evolution, the neutron star
passes through several different evolutionary phases. It is generally 
believed that the population of millisecond pulsars with $P \sim 1-10$ ms 
and $B \sim 10^8-10^9$ G is formed as a result of evolutionary transformations 
in LMXBs. Our calculations show that, apart from millisecond pulsars, one 
more class of pulsars can be formed in LMXBs with a period longer than 
that of millisecond pulsars and with a comparable magnetic field. The only 
difference between these two classes of objects is that the progenitors of 
millisecond pulsars experience accretion with a higher accretion rate 
during the Roche-lobe overflow, whereas accretion onto the progenitors of 
long-period pulsars occurs with a smaller $\dot{M}$. At a smaller 
accretion rate, the neutron star leaves the spin-up line and evolves into 
the region of low magnetic fields and long periods in the $B$-$P$ diagram. 
As a result, it has a relatively long period, $P > 10$ ms, when accretion is 
exhausted. We note that the period of sources, formed by the considered mechanism, 
can  generally be rather long and can reach $\sim 1$ s, but the magnetic field 
should be weak ($\sim 10^8 - 10^{10}$ G) compared to the typical pulsar field.  
The recently discovered accreting pulsar IGR J17480-2446 with $P = 90.6$ ms
and estimated magnetic field $\sim 7 \times 10^8$ G \citep{papitto11}
can be  representative of this class of sources. The period of pulsars in 
this class should satisfy the condition $P > P_{eq}$ for the corresponding 
accretion rate and can vary within a wide range. Perhaps other known 
long-period accreting pulsars, GRO J1744--28 $(P=467$ ms and 
$B \approx 2.4 \times 10^{11}$ G;  \citep{cui} ) and 2A 1822-371 ($P=590$ ms and 
$B \approx 10^{11}$ G; \citep{jonk} ), also belong to this class. 

Our conclusion regarding a possibility of 
IGR J17480-2446 to be a representative of the new class of accreting
pulsars is essentially based on the estimate of the magnetic field
$B \sim 7 \times 10^8$ G made by \cite{papitto11}. This estimate
is provided by means of an additional constraint, the observations of a
815 Hz quasiperiodic oscillations \citep{altamirano11}. 
In this case a model of quasiperiod oscillations is then used to obtain the
innermost disk edge radius and also the magnetic field strength.
Therefore, the estimated value of the magnetic field is model dependent. 
For instance, using the  presence of pulsations throughout the 
considered observation, \cite{papitto11} estimate the magnetic field range 
as $\sim 2 \times 10^8 - 2.4 \times 10^{10}$ G. 
If instead of $7 \times 10^8$ G, one assumes a value of $B \leq
10^{10}$ G from the allowed field range, then IGR J17480-2446 can be 
a normal mildly recycled pulsar. Therefore, more accurate measurements
of the magnetic field are required to confirm that IGR J17480-2446
belongs to the new class. We note, however, that our theoretical modeling 
clearly predicts that there must exist accreting pulsars in LMXB 
that exhibit remarkable departures from the spin-up line in their 
evolution. Even if future measurements  indicate that the
magnetic field of IGR J17480-2446 is sufficiently high, it will be possible to detect representatives 
of this new class  among other accreting pulsars in 
future observations.


\begin{thebibliography}{30}
\expandafter\ifx\csname natexlab\endcsname\relax\def\natexlab#1{#1}\fi

\bibitem[{{Akmal} {et~al.}(1998){Akmal}, {Pandharipande}, \& {Ravenhall}}]{apr}
{Akmal}, A., {Pandharipande}, V.~R., \& {Ravenhall}, D.~G. 1998, \prc, 58, 1804

\bibitem[{{Altamirano} {et~al.}(2010){Altamirano}, {Homan}, {Linares},
  {Patruno}, {Yang}, {Watts}, {Kalamkar}, {Casella}, {Armas-Padilla},
  {Cavecchi}, {Degenaar}, {Russell}, {Kaur}, {van der Klis}, {Rea}, \&
  {Wijnands}}]{altamirano11}
{Altamirano}, D., {Homan}, J., {Linares}, M., {et~al.} 2010, The Astronomer's
  Telegram, 2952, 1

\bibitem[{{Bhattacharya} \& {Srinivasan}(1991)}]{ba91}
{Bhattacharya}, D. \& {Srinivasan}, G. 1991, in {Neutron Stars: Theory and
  Observations}, ed. J.~Ventura \& D.~Pines (Dordrecht: Kluver Academic
  Publisher)

\bibitem[{{Bhattacharya} \& {van den Heuvel}(1991)}]{bhatta91}
{Bhattacharya}, D. \& {van den Heuvel}, E.~P.~J. 1991, \physrep, 203, 1

\bibitem[{{Bonanno} {et~al.}(2014){Bonanno}, {Baldo}, {Burgio}, \&
  {Urpin}}]{boba}
{Bonanno}, A., {Baldo}, M., {Burgio}, G.~F., \& {Urpin}, V. 2014, \aap, 561, L5

\bibitem[{{Bonanno} {et~al.}(2005){Bonanno}, {Urpin}, \&
  {Belvedere}}]{bonanno05}
{Bonanno}, A., {Urpin}, V., \& {Belvedere}, G. 2005, \aap, 440, 199

\bibitem[{{Bonanno} {et~al.}(2006){Bonanno}, {Urpin}, \&
  {Belvedere}}]{bonanno06}
{Bonanno}, A., {Urpin}, V., \& {Belvedere}, G. 2006, \aap, 451, 1049

\bibitem[{{Bondi}(1952)}]{bondi52}
{Bondi}, H. 1952, \mnras, 112, 195

\bibitem[{{Brown} \& {Bildsten}(1998)}]{brown98}
{Brown}, E.~F. \& {Bildsten}, L. 1998, \apj, 496, 915

\bibitem[{{Cui}(1997)}]{cui}
{Cui}, W. 1997, \apjl, 482, L163

\bibitem[{{Fujimoto} {et~al.}(1984){Fujimoto}, {Hanawa}, {Iben}, \&
  {Richardson}}]{fuji84}
{Fujimoto}, M.~Y., {Hanawa}, T., {Iben}, Jr., I., \& {Richardson}, M.~B. 1984,
  \apj, 278, 813

\bibitem[{{Haensel} \& {Zdunik}(1990)}]{hae90}
{Haensel}, P. \& {Zdunik}, J.~L. 1990, \aap, 227, 431

\bibitem[{{Haensel} \& {Zdunik}(2008)}]{zdu08}
{Haensel}, P. \& {Zdunik}, J.~L. 2008, \aap, 480, 459

\bibitem[{{Illarionov} \& {Sunyaev}(1975)}]{Illarionov75}
{Illarionov}, A.~F. \& {Sunyaev}, R.~A. 1975, \aap, 39, 185

\bibitem[{{Jonker} \& {van der Klis}(2001)}]{jonk}
{Jonker}, P.~G. \& {van der Klis}, M. 2001, \apjl, 553, L43

\bibitem[{{Konenkov} \& {Urpin}(1998)}]{konenkov98}
{Konenkov}, D. \& {Urpin}, V. 1998, \mnras, 301, 175

\bibitem[{{Miralda-Escude} {et~al.}(1990){Miralda-Escude}, {Paczynski}, \&
  {Haensel}}]{mira90}
{Miralda-Escude}, J., {Paczynski}, B., \& {Haensel}, P. 1990, \apj, 362, 572

\bibitem[{{Naito} \& {Kojima}(1994)}]{naito94}
{Naito}, T. \& {Kojima}, Y. 1994, \mnras, 266, 597

\bibitem[{{Page} {et~al.}(2004){Page}, {Lattimer}, {Prakash}, \&
  {Steiner}}]{page04}
{Page}, D., {Lattimer}, J.~M., {Prakash}, M., \& {Steiner}, A.~W. 2004, \apjs,
  155, 623

\bibitem[{{Papitto} {et~al.}(2011){Papitto}, {D'A{\`i}}, {Motta}, {Riggio},
  {Burderi}, {di Salvo}, {Belloni}, \& {Iaria}}]{papitto11}
{Papitto}, A., {D'A{\`i}}, A., {Motta}, S., {et~al.} 2011, \aap, 526, L3

\bibitem[{{Potekhin}(1999)}]{pote99}
{Potekhin}, A.~Y. 1999, \aap, 351, 787

\bibitem[{{Pringle} \& {Rees}(1972)}]{pringle72}
{Pringle}, J.~E. \& {Rees}, M.~J. 1972, \aap, 21, 1

\bibitem[{{Shibazaki} {et~al.}(1989){Shibazaki}, {Murakami}, {Shaham}, \&
  {Nomoto}}]{shiba}
{Shibazaki}, N., {Murakami}, T., {Shaham}, J., \& {Nomoto}, K. 1989, \nat, 342,
  656

\bibitem[{{Urpin} {et~al.}(1998{\natexlab{a}}){Urpin}, {Geppert}, \&
  {Konenkov}}]{urpin98mn}
{Urpin}, V., {Geppert}, U., \& {Konenkov}, D. 1998{\natexlab{a}}, \mnras, 295,
  907

\bibitem[{{Urpin} {et~al.}(1998{\natexlab{b}}){Urpin}, {Geppert}, \&
  {Konenkov}}]{urpin98aa}
{Urpin}, V., {Geppert}, U., \& {Konenkov}, D. 1998{\natexlab{b}}, \aap, 331,
  244

\bibitem[{{Urpin} {et~al.}(1998{\natexlab{c}}){Urpin}, {Konenkov}, \&
  {Geppert}}]{urpin98hm}
{Urpin}, V., {Konenkov}, D., \& {Geppert}, U. 1998{\natexlab{c}}, \mnras, 299,
  73

\bibitem[{{Urpin} {et~al.}(1994){Urpin}, {Chanmugam}, \& {Sang}}]{urpin94}
{Urpin}, V.~A., {Chanmugam}, G., \& {Sang}, Y. 1994, \apj, 433, 780

\bibitem[{{Wendell} {et~al.}(1987){Wendell}, {van Horn}, \&
  {Sargent}}]{wendell87}
{Wendell}, C.~E., {van Horn}, H.~M., \& {Sargent}, D. 1987, \apj, 313, 284

\bibitem[{{Wijers}(1997)}]{wije}
{Wijers}, R.~A.~M.~J. 1997, \mnras, 287, 607

\bibitem[{{Zdunik} {et~al.}(1992){Zdunik}, {Haensel}, {Paczynski}, \&
  {Miralda-Escude}}]{zdu92}
{Zdunik}, J.~L., {Haensel}, P., {Paczynski}, B., \& {Miralda-Escude}, J. 1992,
  \apj, 384, 129

\end{thebibliography}

\end{document}